\documentclass[]{IEEEapm}

\jmonth{September}
\pubyear{2026}

\usepackage{amsmath,amssymb,amsfonts}
\usepackage{array}
\usepackage[caption=false,font=normalsize,labelfont=sf,textfont=sf]{subfig}
\usepackage{textcomp}
\usepackage{stfloats}
\usepackage{verbatim}
\usepackage{graphicx}
\usepackage{cite}
\usepackage{tabularx}
\usepackage{booktabs}

\def\BibTeX{{\rm B\kern-.05em{\sc i\kern-.025em b}\kern-.08em
    T\kern-.1667em\lower.7ex\hbox{E}\kern-.125emX}}

\usepackage{acronym}
\acrodef{5G}{fifth generation}
\acrodef{B5G}{beyond fifth generation}
\acrodef{MIMO}{multiple-input multiple-output}
\acrodef{SISO}{single-input single-output}
\acrodef{RF}{radio frequency}
\acrodef{LoS}{line-of-sight}
\acrodef{NLoS}{non-line-of-sight}
\acrodef{AoA}{angle-of-arrival}
\acrodef{AoD}{angle-of-departure}
\acrodef{UPA}{uniform planar array}
\acrodef{ARV}{array response vector}
\acrodef{EM}{electromagnetic}
\acrodef{MA}{movable antenna}
\acrodef{BS}{base station}
\acrodef{UE}{user equipment}
\acrodef{RA}{reconfigurable antenna}
\acrodef{ERA}{element-reconfigurable array}
\acrodef{SRA}{spatially reconfigurable antenna}
\acrodef{AWGN}{additive white Gaussian noise}
\acrodef{HFSS}{High-Frequency Structure Simulator}
\acrodef{WMMSE}{weighted minimum mean-square error}
\acrodef{BCD}{block coordinate descent}
\acrodef{ZF}{zero-forcing}
\acrodef{6G}{sixth generation}
\acrodef{3D}{three-dimensional}
\acrodef{2D}{two-dimensional}
\acrodef{ISAC}{integrated sensing and communications}
\acrodef{BER}{bit error rate}
\acrodef{RIS}{reconfigurable intelligent surface}
\acrodef{CRB}{Cramér–Rao bound}
\acrodef{MSE}{mean-squared error}
\acrodef{RMSE}{root mean-squared error}
\acrodef{SNR}{signal-to-noise ratio}
\acrodef{SINR}{signal-to-interference-plus-noise ratio}
\acrodef{TOA}{time-of-arrival}
\acrodef{TDOA}{time-difference-of-arrival}
\acrodef{SDR}{software-defined radio}
\acrodef{DMA}{dynamic metasurface antenna}
\acrodef{SIM}{stacked intelligent metasurface}
\acrodef{PRA}{pattern-reconfigurable antenna}
\acrodef{FRA}{frequency-reconfigurable antenna}
\acrodef{PoRA}{polarization-reconfigurable antenna}
\acrodef{PIN}{positive–intrinsic–negative}
\acrodef{DAC}{digital-to-analog converter}
\acrodef{CSI}{channel state information}
\acrodef{UAV}{unmanned aerial vehicle}
\acrodef{3GPP}{3rd Generation Partnership Project}
\acrodef{RAN}{radio access network}
\acrodef{KPI}{key performance indicator}

\usepackage{color}

\usepackage{tikz} 
\usepackage[utf8]{inputenc}
\usepackage{pgfplots}
\usepackage{tabu,longtable}
\usepackage{diagbox}
\usepackage{threeparttable}
\usepackage{makecell}
\usepackage{multirow} 
\usepackage{units} 		
\usepackage{bm} 		
\usepackage{amssymb} 	

\usepackage{hyperref} 	
\usepackage{amsmath}

\usepackage{algorithm}
\usepackage{algpseudocode}

\makeatletter
\algnewcommand{\LineComment}[1]{\Statex \hskip\ALG@thistlm \(\triangleright\) #1}
\makeatother
\usepackage{pgfplots}
\usepackage{tikz}
\usetikzlibrary{calc}
\makeatletter
\newcommand{\gettikzxy}[3]{%
  \tikz@scan@one@point\pgfutil@firstofone#1\relax
  \edef#2{\the\pgf@x}%
  \edef#3{\the\pgf@y}%
}

\begin{document}
\bstctlcite{IEEEexample:BSTcontrol} 

\title{Reconfigurable Antennas {for Next-Generation Wireless Communications}: Technologies, Prototypes, Architectures, and Signal Processing}

\author{Pinjun~Zheng\IEEEauthorrefmark{2},~\IEEEmembership{Member,~IEEE}, Ruiqi~Wang\IEEEauthorrefmark{1}, Yuchen~Zhang\IEEEauthorrefmark{1},~\IEEEmembership{Member,~IEEE},  
Md.~Jahangir~Hossain\IEEEauthorrefmark{2},~\IEEEmembership{Senior~Member,~IEEE}, Anas~Chaaban\IEEEauthorrefmark{2},~\IEEEmembership{Senior~Member,~IEEE}\\ 
Atif~Shamim\IEEEauthorrefmark{1},~\IEEEmembership{Fellow,~IEEE}, Tareq~Y.~Al-Naffouri\IEEEauthorrefmark{1},~\IEEEmembership{Fellow,~IEEE}}

\affil{\IEEEauthorrefmark{2}School of Engineering, the University of British Columbia, Kelowna, BC, V1V 1V7, Canada\\
\IEEEauthorrefmark{1}CEMSE, King Abdullah University of Science and Technology, Thuwal, 23955-6900, Saudi Arabia \vspace{-4em}}

\maketitle

\markboth{IEEE Antennas and Propagation Magazine}{An IEEE Antennas and Propagation Society Publication}

\begin{receivedinfo}%
\end{receivedinfo}

\begin{abstract}
The transition to sixth-generation (6G) networks calls for wireless transceivers with enhanced adaptability and efficiency. Reconfigurable antennas (RAs) have emerged as a promising solution, enabling dynamic control over the electromagnetic properties of individual antenna elements. Their integration into antenna arrays is particularly attractive due to their adaptability and potential for improved energy efficiency. This article provides a comprehensive overview of RA technologies for advanced communication systems, encompassing hardware advancements, early experimental studies, novel system architectures, and key signal processing challenges. From the combined perspective of antenna design and communication operation, we highlight the potential of RAs to enable next-generation wireless communications, while also identifying key challenges and promising research opportunities. 
\end{abstract}


\section{Introduction}

As modern wireless communication systems advance toward the \ac{B5G} and \ac{6G} eras, using high-frequency bands and large-scale antenna arrays has become a prominent trend to meet the ever-growing demand for capacity and reliability. {Although the actual deployment standard for \ac{6G} has not yet been finalized, the deployment scenarios for evaluating \ac{6G} \ac{RAN} performance have been outlined in the \ac{3GPP} TR 38.914~\cite{3gpp38914}. As summarized in Table~\ref{tab:3GPP}, \ac{6G} \acp{BS} are expected to deploy up to thousands of antenna elements when operating above \unit[7]{GHz}. However, scaling up antenna arrays introduces substantial challenges in hardware complexity, deployment cost, and energy consumption.}

 A critical lesson learned from the \ac{5G} networks is that next-generation wireless systems must not only push performance boundaries but also prioritize energy and cost efficiency~\cite{Andrews20246G}. As shown in Table~\ref{tab:EE}, 6G targets a $100\times$ improvement in energy efficiency, meaning it is expected to transmit 100 times more data for the same amount of energy. Achieving this goal requires both more energy-efficient transceiver designs and a reduction in the absolute power consumption of the \ac{BS}. Samsung’s evaluation indicates that a 6G architecture building on context-aware and on-demand signaling may reduce power consumption by more than 30\% compared with 5G~\cite{Samsung2025EnergySaving}; however, this remains far from the ambitious $100\times$ goal in energy efficiency, highlighting the need for more efficient transceiver designs. Ultimately, achieving higher communication rates under constrained hardware, bandwidth, and power budgets relies on utilizing all available resources more efficiently. A key aspect is the precise delivery of energy to the intended receivers while minimizing spatial leakage and interference. 

\Acp{RA} present a promising solution by introducing new control dimensions at the antenna level. This added flexibility can be leveraged to improve system adaptability and power efficiency. Recently, interest in \acp{RA} has been rising rapidly within the communications community. A growing body of work shows that \ac{MIMO} systems equipped with RA arrays can achieve substantial gains in various communication \acp{KPI} over systems using conventional static antennas~\cite{Shlezinger2021Dynamic,Rasilainen2023Hardware,Ying2024Reconfigurable,Castellanos2025Embracing,New2025Fluid,Zhang2025Compact,Zheng2025Tri,Deshpande2026Signal}. Despite decades of extensive research on \acp{RA} within the antenna community, their application in communication systems, particularly for emerging \ac{6G}-era wireless systems, remains an active area of exploration. This article aims to review \ac{RA} technologies, present early experimental studies, discuss emerging transceiver architectures, and identify key signal processing challenges.

\begin{table}[!t]
\centering
\scriptsize
\caption{Deployment Scenarios for 6G RAN Evaluation in 3GPP TR 38.914~\cite{3gpp38914}}
\label{tab:3GPP}
\begin{IEEEeqnarraybox}[\IEEEeqnarraystrutmode\IEEEeqnarraystrutsizeadd{2pt}{1pt}]{v/c/v/c/v/c/v}
\IEEEeqnarrayrulerow\\
&\mbox{Carrier Frequency} && \mbox{System Bandwidth} && \mbox{BS Antenna Elements} &\\
\IEEEeqnarraydblrulerow\\
\IEEEeqnarrayseprow[3pt]\\
&\mbox{700 MHz} && \mbox{Up to 60 MHz} && \mbox{Up to 64} &\\
\IEEEeqnarrayrulerow\\
\IEEEeqnarrayseprow[3pt]\\
&\mbox{2 GHz} && \mbox{Up to 200 MHz} && \mbox{Up to 288} &\\
\IEEEeqnarrayrulerow\\
\IEEEeqnarrayseprow[3pt]\\
&\mbox{4 GHz} && \mbox{Up to 300 MHz} && \mbox{Up to 576} &\\
\IEEEeqnarrayrulerow\\
\IEEEeqnarrayseprow[3pt]\\
&\mbox{7 GHz} && \mbox{Up to 400 MHz} && \mbox{Up to 2304} &\\
\IEEEeqnarrayrulerow\\
\IEEEeqnarrayseprow[3pt]\\
&\mbox{15 GHz} && \mbox{Up to 400 MHz} && \mbox{Up to 2304} &\\
\IEEEeqnarrayrulerow\\
\IEEEeqnarrayseprow[3pt]\\
&\mbox{30 GHz} && \mbox{Up to 1 GHz} && \mbox{Up to 4096} &\\
\IEEEeqnarrayseprow[3pt]\\
\IEEEeqnarrayrulerow
\end{IEEEeqnarraybox}
\end{table}

\begin{table}[!t]
\centering
\scriptsize
\caption{Targeted Energy-Saving Gains in 6G}
\label{tab:EE}
\begin{IEEEeqnarraybox}[\IEEEeqnarraystrutmode\IEEEeqnarraystrutsizeadd{2pt}{1pt}]{v/c/v/c/v/c/v}
\IEEEeqnarrayrulerow\\
&\mbox{KPI} && \mbox{5G Baseline} && \mbox{6G Target} &\\
\IEEEeqnarraydblrulerow\\
\IEEEeqnarrayseprow[3pt]\\
&\mbox{Energy Efficiency} && \unit[10^7]{bits/J}\mbox{~\cite{Wang2023On}} && \unit[10^9]{bits/J}\mbox{~\cite{Wang2023On,Ahmadi2025Toward}} &\\
\IEEEeqnarrayrulerow\\
\IEEEeqnarrayseprow[3pt]\\
&\mbox{BS Energy Consumption} && \unit[0.7 - 4.5]{kW}\mbox{~\cite{Ma2024Sustainable}} && \mbox{Reduce by }30\% \mbox{ or more~\cite{Samsung2025EnergySaving}}&\\
\IEEEeqnarrayseprow[3pt]\\
\IEEEeqnarrayrulerow
\end{IEEEeqnarraybox}
\end{table}

\section{Reconfigurable Antenna Hardware}\label{sec:Hardware}

\Acp{RA} provide an important hardware pathway toward adaptive wireless systems by enabling dynamic control of antenna characteristics. Compared with conventional fixed antennas, such flexibility is attractive for wireless systems operating under dynamic propagation conditions, wide coverage requirements, and heterogeneous deployment scenarios. This section reviews \ac{RA} hardware from both antenna-design and deployment-oriented perspectives, highlighting major functional types, practical design tradeoffs, industrial progress, and key insights for future adoption.

\subsection{Functional Types of Reconfigurable Antennas}

RAs are particularly relevant to 5G and 6G systems because high-frequency wireless links face severe blockage, beam misalignment, polarization mismatch, limited device aperture, and power-consumption constraints. These challenges are difficult to solve through baseband or \ac{RF} processing alone. By reconfiguring radiation pattern, operating frequency, or polarization at the antenna level, RAs provide an additional hardware degree of freedom to improve coverage, robustness, and energy efficiency.

\begin{figure*}[t]
  \centering
  \includegraphics[width=\linewidth]{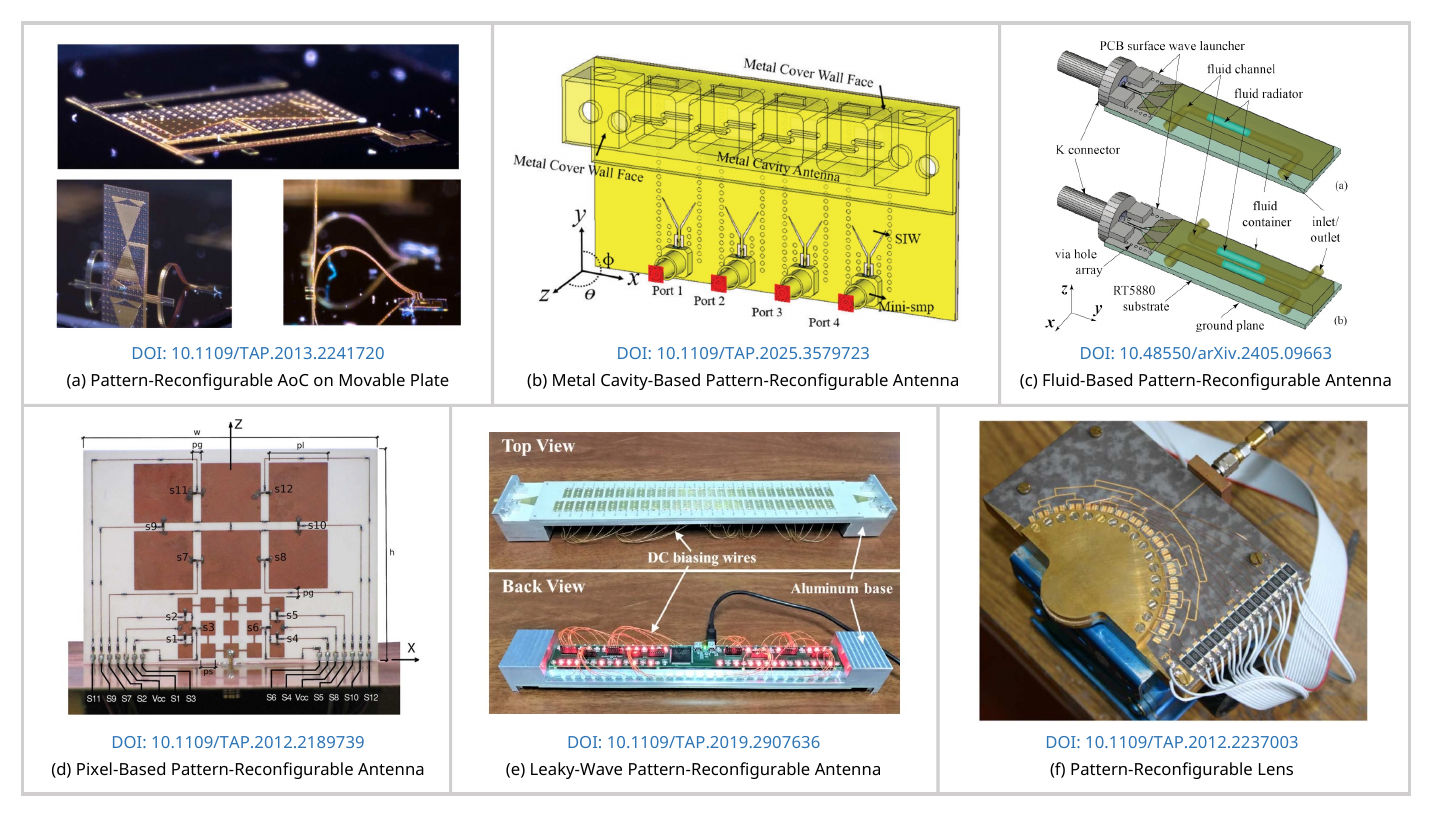}
  \vspace{-1.8em}
  \caption{ 
      Exemplary prototypes of PRA hardware. (a)~Antenna-on-Chip on movable plate. (b)~Metal cavity-based PRA. (c)~Fluid-based PRA. (d)~Pixel-based PRA. (e)~Leaky-wave PRA. (f)~Pattern-reconfigurable lens. The images are taken from the references corresponding to the DOIs provided in the figure.
    }
  \label{fig_RQ1}
\end{figure*}

\subsubsection{Pattern-Reconfigurable Antenna}

One of the most common types of \ac{RA} is the \emph{\ac{PRA}}, which dynamically modifies the radiation pattern of the antenna. As illustrated in Fig.~\ref{fig_RQ1}, pattern reconfigurability can be realized through different antenna technologies, including movable radiators, phased arrays, reconfigurable reflectarrays, reconfigurable cavities, fluid-based structures, pixel-based antennas, leaky-wave antennas, and reconfigurable lenses~\cite{Marnat2013AoC,Zhang2025PhasedArray,Shen2024FAS,Rodrigo2012Pixel,Shen2017TAP,Shen2018TAP,Li2019Leaky,Lafond2013Lens}. These designs enable the antenna to switch among broadside, endfire, oblique, or multi-beam radiation states, thereby providing additional spatial degrees of freedom. The main advantage of \acp{PRA} is their ability to adapt the radiation behavior of a single antenna to different coverage requirements without relying on array-factor control; as will become clear later, combining such single-antenna pattern reconfigurability with array-factor control can enable highly flexible beampattern synthesis. By switching among different radiation states, a \ac{PRA} can redirect the main beam, reshape the coverage region, or suppress radiation toward undesired directions. This feature is particularly attractive for applications requiring environment-dependent coverage adaptation. However, the main challenge of \acp{PRA} lies in maintaining stable antenna performance across different radiation states. Since pattern reconfiguration usually modifies the current distribution and modal behavior of the antenna, different states may exhibit nonuniform gain, beamwidth, sidelobe level, impedance matching, radiation efficiency, and polarization purity. Therefore, a practical \ac{PRA} should not only provide diverse radiation patterns, but also preserve consistent radiation quality, compact element size, and array-compatible control complexity across all states.

\begin{figure*}[t]
  \centering
  \includegraphics[width=\linewidth]{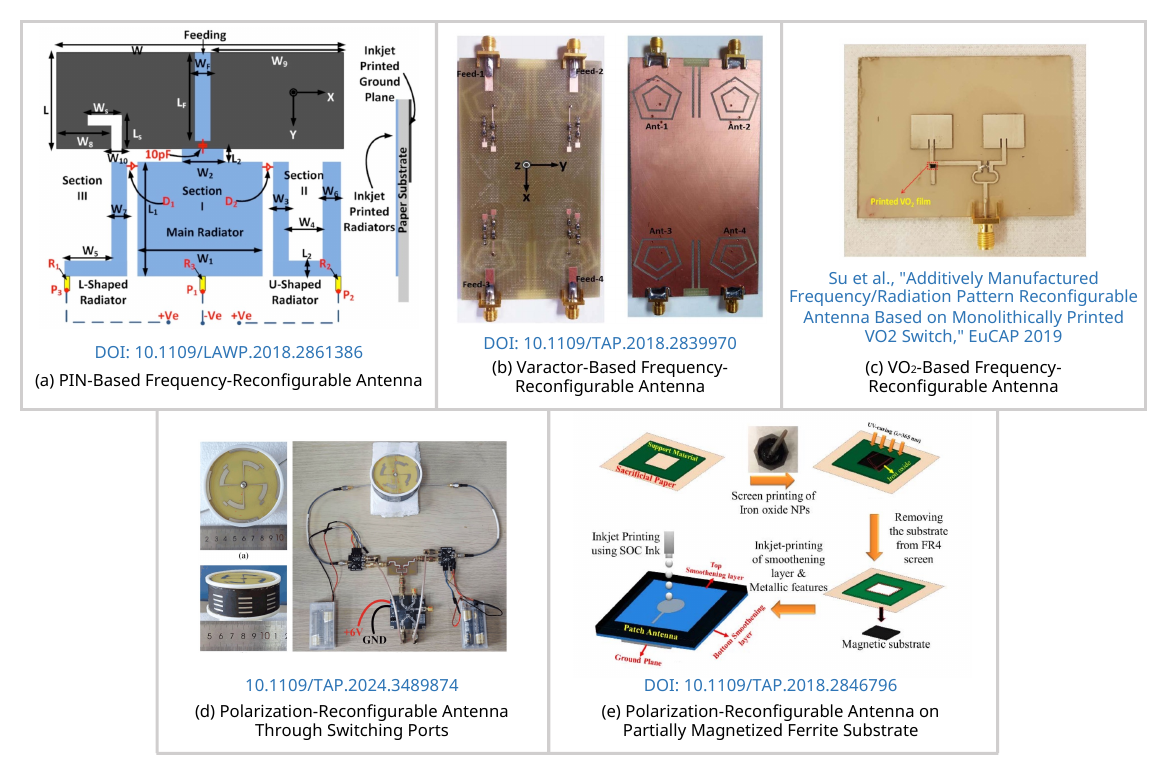}
  \vspace{-1.8em}
  \caption{ 
      Exemplary prototypes of FRAs and PoRAs. (a)~FRA designed with PIN diodes. (b)~FRA designed with varactor diodes. (c)~FRA designed with phase-changing switches. (d)~PoRA through port switching. (e)~PoRA on magnetized substrate. The images are taken from the references corresponding to the DOIs provided in the figure.
    }
  \label{fig_RQ2}
\end{figure*}

\subsubsection{Frequency-Reconfigurable Antenna}

Another representative type is the \emph{\ac{FRA}}, which adjusts the operating frequency of the antenna. Since the resonant frequency of an antenna is closely related to its effective electrical length and equivalent reactive loading~\cite{Songnan2009FRA}, frequency reconfigurability can be achieved by modifying the current path or tuning the effective capacitance and inductance of the antenna. Representative examples are shown in Fig.~\ref{fig_RQ2}(a)--(c), including PIN-diode-based, varactor-based, and phase-change-material-based implementations~\cite{Abutarboush2018AWPL,Hussain2018FRA,Singh2023PhaseChange,Su2019EuCAP}. \Acp{FRA} are useful for multi-band operation, spectrum agility, and compact wireless platforms where fixed antennas are difficult to accommodate. Nevertheless, frequency reconfiguration is not merely a matter of shifting the resonance. A wide tuning range is often accompanied by reduced radiation efficiency, narrower bandwidth, increased loss, or degraded linearity, especially when tunable components are introduced into high-current regions. In addition, the radiation pattern may vary across the tuning range because different resonant modes or current distributions can dominate at different frequencies. Therefore, a practical \ac{FRA} should maintain stable radiation behavior, acceptable impedance matching, and low loss over all target frequency states.

\subsubsection{Polarization-Reconfigurable Antenna}

The third important category is the \emph{\ac{PoRA}}, which reconfigures the polarization state of the antenna. As shown in Fig.~\ref{fig_RQ2}(d) and Fig.~\ref{fig_RQ2}(e), polarization reconfiguration can be realized by changing the excited mode, feeding condition, current path, or material state~\cite{Ghaffar2018Ferrite}. This capability allows the antenna to switch among linear polarization, left-hand circular polarization (LHCP), and right-hand circular polarization (RHCP), thereby improving polarization diversity, link robustness, and tolerance to user orientation or propagation uncertainty. The design challenge of \acp{PoRA} lies in maintaining polarization quality across multiple states. For circularly polarized antennas, the axial ratio, gain, impedance matching, and radiation direction need to be simultaneously controlled. Switching the polarization state may also change the current distribution and disturb the radiation pattern or operating bandwidth. As a result, the number of available polarization states alone is not sufficient to evaluate a \ac{PoRA}. More importantly, different polarization states should provide comparable gain, stable beam direction, acceptable axial-ratio bandwidth, and consistent radiation efficiency.

\subsection{Industrial Progress and Practical Adoption}

From an industrial point of view, \ac{RA} technologies have already moved beyond laboratory demonstrations and have been commercialized in wireless products. In practical mobile terminals, such reconfigurability has been widely adopted at the \ac{RF} front-end level, where antenna management circuits use tuning, swapping, and coupling functions to select the best-performing antenna or adjust the aperture and impedance of the antenna for different frequency bands and user conditions~\cite{Skyworks_Mobile_Antenna_Management,Qorvo_Antenna_Control}. 

Beyond mobile terminals, \ac{RA} technologies have also entered other commercial wireless platforms. For example, Kymeta has commercialized electronically steered flat-panel satellite terminals for mobile connectivity, RUCKUS Networks has deployed the BeamFlex hardware-based adaptive antenna system in enterprise Wi-Fi access points, KYOCERA AVX has developed active-steering smart antenna modules for real-time radiation-pattern optimization, and Ignion, together with STMicroelectronics, has promoted virtual or tunable antenna solutions for compact IoT devices such as smart meters and asset trackers~\cite{Kymeta2023HawkU8,RUCKUSBeamFlex,KYOCERAAVXActiveSteering,STIgnion2025}.

\subsection{Deployment-Oriented Design Considerations}

The design requirements of \acp{RA} depend strongly on their deployment scenarios, particularly when they are considered for 6G systems with higher carrier frequencies, denser antenna integration, distributed access points, and more dynamic propagation environments. 

For \ac{BS}-side deployment, RAs face a fundamental tradeoff between reconfigurability and array scalability. Strong reconfigurability often requires larger antenna apertures, additional tuning components, or more complex biasing networks, which can limit array scalability. In large-scale arrays, state-dependent variations in element gain, phase center, impedance, and polarization can accumulate across many elements, distort the array manifold, increase the calibration burden, and degrade beamforming performance. Accordingly, \acp{RA} for future \acp{BS} should be evaluated not only in terms of \ac{EM} flexibility, but also in terms of their ability to support efficient and rapidly reconfigurable array deployment.

For user-side deployment, the design constraints are different from BS-side deployment because handheld, wearable~\cite{Ruiqi2023RFID}, and vehicular platforms~\cite{Zhan2022Vehicular} impose stricter constraints on size, form factor, power consumption, thermal budget, and integration space, which become more stringent at the high-frequency bands envisioned for 6G. These constraints have always existed for user devices, but they become more challenging for RAs because reconfigurable elements require additional switches, tunable materials, control lines, and bias circuits within an already compact platform. Limited antenna aperture and low-profile packaging constrain reconfiguration complexity, while user proximity and body loading introduce strong and dynamic perturbations to impedance matching, radiation characteristics, and efficiency. As a result, reconfigurability at the user side is often exploited to enhance robustness against blockage, detuning, and interference, rather than to realize highly flexible beamforming. Practical user-side \acp{RA} should also prioritize low loss, fast adaptation, compliance with specific absorption rate (SAR) safety limits, and stable performance under near-field coupling.

\subsection{Summary and Key Insights}

The \acp{RA} discussed above are categorized according to the primary \ac{EM} functions through various reconfiguration technologies, which exhibit distinct characteristics in terms of tuning range, switching speed, control complexity, element size, and array compatibility. Detailed comparisons have been conducted in the literature, and we refer the reader to, e.g., Table~1 and Table~3 in~\cite{ojaroudi2020reconfigurable} and Table~2 in~\cite{Haupt2013reconfigurable} for comprehensive summaries across key design dimensions. For completeness, we also provide a more focused comparison in Table~\ref{ComparisonTable}, which compares several representative \ac{RA} designs in terms of reconfiguration mechanism, element size, switching speed, control complexity, and array compatibility. {For clarity, the qualitative metrics in Table~\ref{ComparisonTable} are defined as follows. Switching speed is classified as Fast when reconfiguration is achieved electronically without macroscopic physical movement, and as Slow when mechanical or fluidic motion is required. Control complexity is classified as Low for simple local control with minimal auxiliary circuitry, Medium for multiple switching/tuning controls or moderate actuation requirements, and High for distributed control or external actuation hardware with relatively high implementation complexity. Array compatibility is evaluated jointly based on the antenna-element footprint and the feasibility of integrating the associated reconfiguration mechanism into compact arrays: High denotes designs readily scalable to dense array integration, Medium denotes moderate constraints in element size or control implementation, and Low denotes significant limitations due to a large footprint or bulky/complex actuation hardware.}

Although numerous \ac{RA} hardware designs have been reported, many of them demonstrate reconfigurability primarily at the single-element level rather than at the array level. A key reason is that many \ac{RA} structures exhibit relatively large physical dimensions, often exceeding half a wavelength, as shown in Table~\ref{ComparisonTable}. This size constraint complicates their integration into dense large-scale arrays and may exacerbate grating-lobe effects. Addressing this challenge requires \ac{RA} designs that are inherently compatible with large-scale array integration while still supporting element-level reconfigurability.

Furthermore, several practical challenges need to be considered in real-world systems. Reconfiguration mechanisms inevitably introduce loss, finite tuning resolution, parasitic effects, and hardware nonlinearity, which degrade radiation efficiency and distort the transmitted signals, especially at high frequencies~\cite{pozar2021microwave,Kolomvakis2025Nonlinear}. In addition, dynamic reconfiguration may cause intermodulation distortion and spectral regrowth~\cite{Luo2024Modeling}. These non-idealities are often neglected in idealized models but fundamentally limit achievable performance gains. Therefore, future \ac{RA} research should explicitly incorporate hardware-aware modeling of practical nonlinear behavior to enable reliable performance evaluation and practical deployment.

\begin{table*}[t]
  \renewcommand{\arraystretch}{1.4}
  \begin{center}
  \caption{Comparison of Existing Reconfigurable Antenna Systems}
  \label{ComparisonTable}
  \scriptsize

  \resizebox{\textwidth}{!}{
  \begin{tabular}{
    !{\vrule width1pt} >{\centering\arraybackslash}p{0.5cm} !{\vrule width1pt}
    >{\centering\arraybackslash}p{3.5cm} !{\vrule width1pt}
    >{\centering\arraybackslash}p{2.7cm} !{\vrule width1pt}
    >{\centering\arraybackslash}p{1.7cm} !{\vrule width1pt}
    >{\centering\arraybackslash}p{3cm} !{\vrule width1pt}
    >{\centering\arraybackslash}p{1.2cm} !{\vrule width1pt}
    >{\centering\arraybackslash}p{1.3cm} !{\vrule width1pt}
    >{\centering\arraybackslash}p{1.3cm} !{\vrule width1pt}
  }

    \Xhline{1pt}
    Ref. &
    Type \& Technique &
    Reconfigurable Mechanism & 
    Antenna Element Size ($\lambda \times \lambda$) &
    Tuning Range &
    Switching Speed &
    Control Complexity &
    Array Compatibility \\
    \hline
    \hline

    \cite{Marnat2013AoC} &
    On-chip Antenna; Movable plate &
    MEMS-based movable plate &
    0.59 $\times$ 0.51 &  
    Limited (two discrete reconfiguration states) &
    Slow &
    Medium &
    High \\

    \hline

    \cite{Zhang2025PhasedArray} &
    Metal cavity; Multimode resonances &
    Mechanically changing the metal cover wall &
    0.93 $\times$ 2.73 &
    Limited (three discrete beam states)  &
    Slow &
    Medium &
    High \\

    \hline

    \cite{Shen2024FAS} &
    Liquid-based; Surface-wave fluid antenna &
    Micropump through silicon tubes &
    2.86 $\times$ 0.86 &
    High (continuous position reconfiguration) &
    Slow &
    High &
    Low \\

    \hline

    \cite{Rodrigo2012Pixel} &
    Pixel-based; Multiple-sized pixels &
    PIN diode &
    0.735 $\times$ 0.53 &
    Moderate &
    Fast &
    Medium &
    Medium \\

    \hline

    \cite{Hussain2018FRA} &
    Slot-based; MIMO antenna &
    Varactor diode &
    0.65 $\times$ 0.33 &
    High (continuous frequency tuning) &
    Fast &
    Medium &
    High \\

    \hline

    \cite{Lafond2013Lens} &
    Lens antenna; Luneburg lens with multisources &
    MMIC switch &
    5.6 $\times$ 5.6 &
    Moderate &
    Fast &
    High &
    Low \\

    \hline
    
    {\cite{Zhan2025TAP}} &
    Dielectric resonator antenna; Odd/Even mode &
    PIN diode &
    0.65 $\times$ 0.81 &
    Moderate &
    Fast &
    Medium &
    Medium \\

    \hline
    
    \cite{Wang2025Electromagnetically} &
    Liquid-based; Directors and reflectors with planar monopole &
    Liquid metal controlled by pumping machine &
    0.5 $\times$ 0.5 &
    High &
    Slow &
    High &
    High \\

    \Xhline{1pt}
  \end{tabular}
  }
  \end{center}
\end{table*}

\section{Array Integration for Reconfigurable Antennas}

{Building on reconfigurable antenna technologies, this section introduces the concept of \ac{ERA}. Antenna arrays are an essential technology for wireless communication systems, as highlighted in Table~\ref{tab:3GPP}. Conventional arrays shape their beampatterns by manipulating the array factor, i.e., by adjusting the excitation amplitudes and phases of individual elements, while each element’s radiation pattern remains fixed. In contrast, by using \acp{RA}, ERAs incorporate reconfigurability within each radiating element in addition to array-factor control. This added degree of freedom enables more adaptive and energy-efficient beamforming. Such advantages are especially valuable for future communication systems, where they can enhance communication rate and coverage under stringent power constraints~\cite{Liu2025Reconfigurable,Emadi20256G}.}

\subsection{Element-Reconfigurable Array Hardware}

\begin{figure*}[t]
  \centering
  \includegraphics[width=0.9\linewidth]{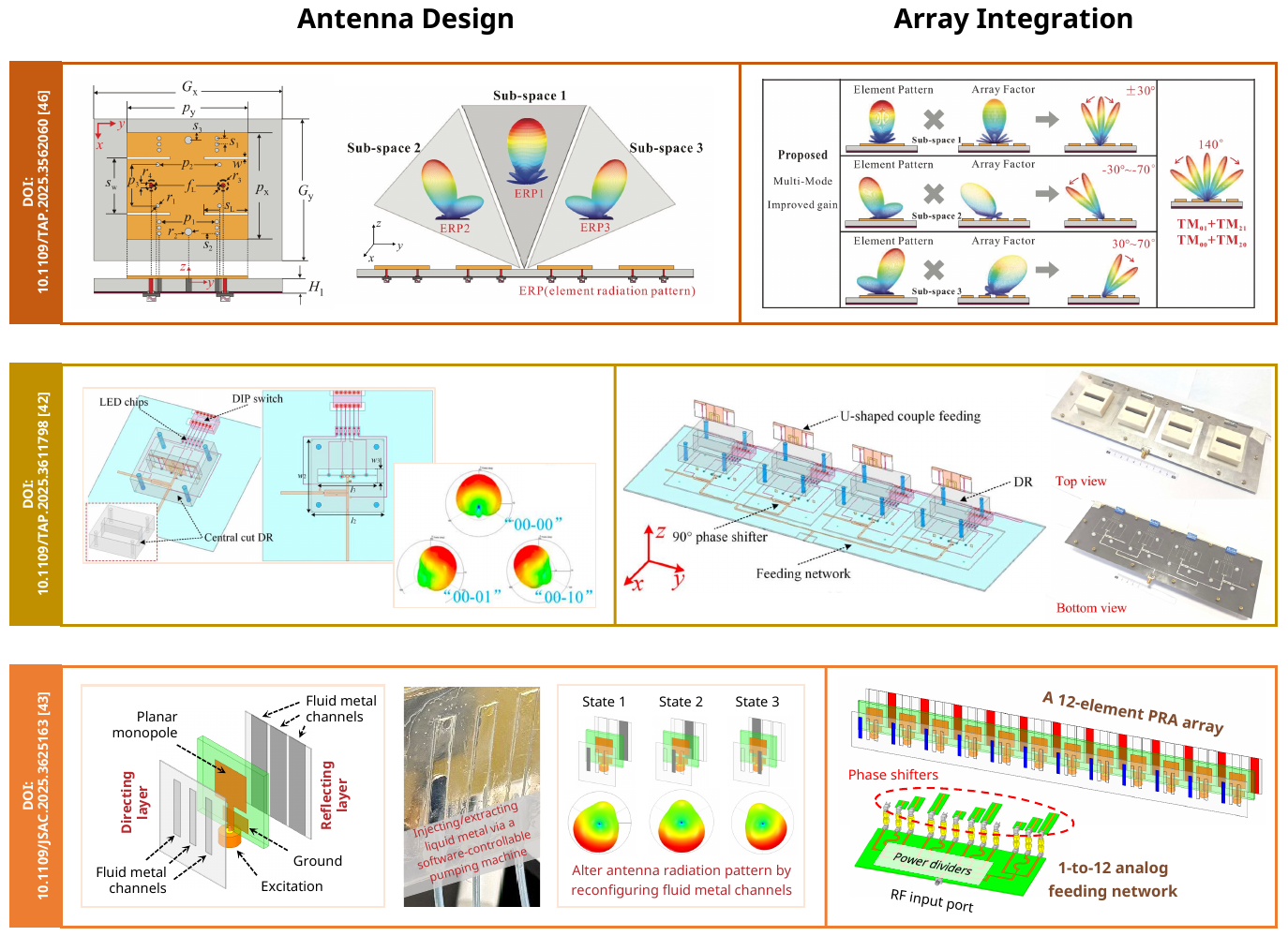}
\caption{
{Examples of array designs exploiting element-level radiation-pattern diversity. 
The designs in~\cite{Zhan2025TAP,Wang2025Electromagnetically} are ERAs with dynamically reconfigurable elements, whereas~\cite{Yan2025TAP} is a static element-pattern-engineered phased-array reference.}
}
\label{fig_ERA_Concept}
\end{figure*} 

{Figure~\ref{fig_ERA_Concept} presents a related static element-pattern-engineered array in~\cite{Yan2025TAP} and two recently reported \ac{ERA} hardware designs in~\cite{Zhan2025TAP,Wang2025Electromagnetically}.} Specifically,~\cite{Yan2025TAP} demonstrates a design in which four TM modes of a microstrip patch antenna are utilized to realize both broadside and dual-beam radiation at the same frequency through different shorting pin loading configurations. Although this design is static without reconfigurable components, distinct radiation patterns are achieved in different elements by tailoring the geometrical configurations, and an array constructed with such geometrically varied elements can achieve wide-angle beamforming up to $70^\circ$. {Therefore,~\cite{Yan2025TAP} is included here as a static reference to illustrate the potential of element-pattern engineering for wide-angle beamforming, rather than as a genuine ERA.} In~\cite{Zhan2025TAP}, an ERA element based on a U-shaped feeding dielectric resonator antenna is designed using even-odd mode analysis to realize three-state radiation patterns. Consequently, the overall array achieves cost-effective wide-angle beam scanning with low sidelobe levels. Nevertheless, the relatively large element size of 0.65$\lambda$ $\times$ 0.81$\lambda$ introduces potential grating lobe issues. In~\cite{Wang2025Electromagnetically}, a liquid-metal-based ERA prototype is proposed, where fluidic director and reflector structures are coupled with a planar monopole antenna. A total of 64 radiation states are achieved with a compact element size of 0.5$\lambda$ $\times$ 0.5$\lambda$, making it suitable for large-scale array implementations.

The aforementioned designs focus on ERAs with radiation pattern reconfigurability at the element level, i.e., \ac{PRA} arrays. In addition to that, arrays of \acp{FRA} and \acp{PoRA} have also begun to be investigated~\cite{Castellanos2024TWC,Liu2025Reconfigurable,Lee2026Integrated}. Nevertheless, these approaches are still at an early research stage and lack validation through practical prototypes and experimental results.

\subsection{An Experimental Study of Wireless Communications with ERAs}

To provide further insights into the advantages of ERAs in wireless communications, we now present a communication experiment using the ERA prototype in~\cite{Wang2025Electromagnetically}.\footnote{Portions of these results have previously appeared in~\cite{Wang2025Electromagnetically}.} This prototype operates at 3.55~GHz, which is the center frequency of the \ac{5G} n78 band.

\begin{figure*}[t]
\centering
\includegraphics[width=0.95\linewidth]{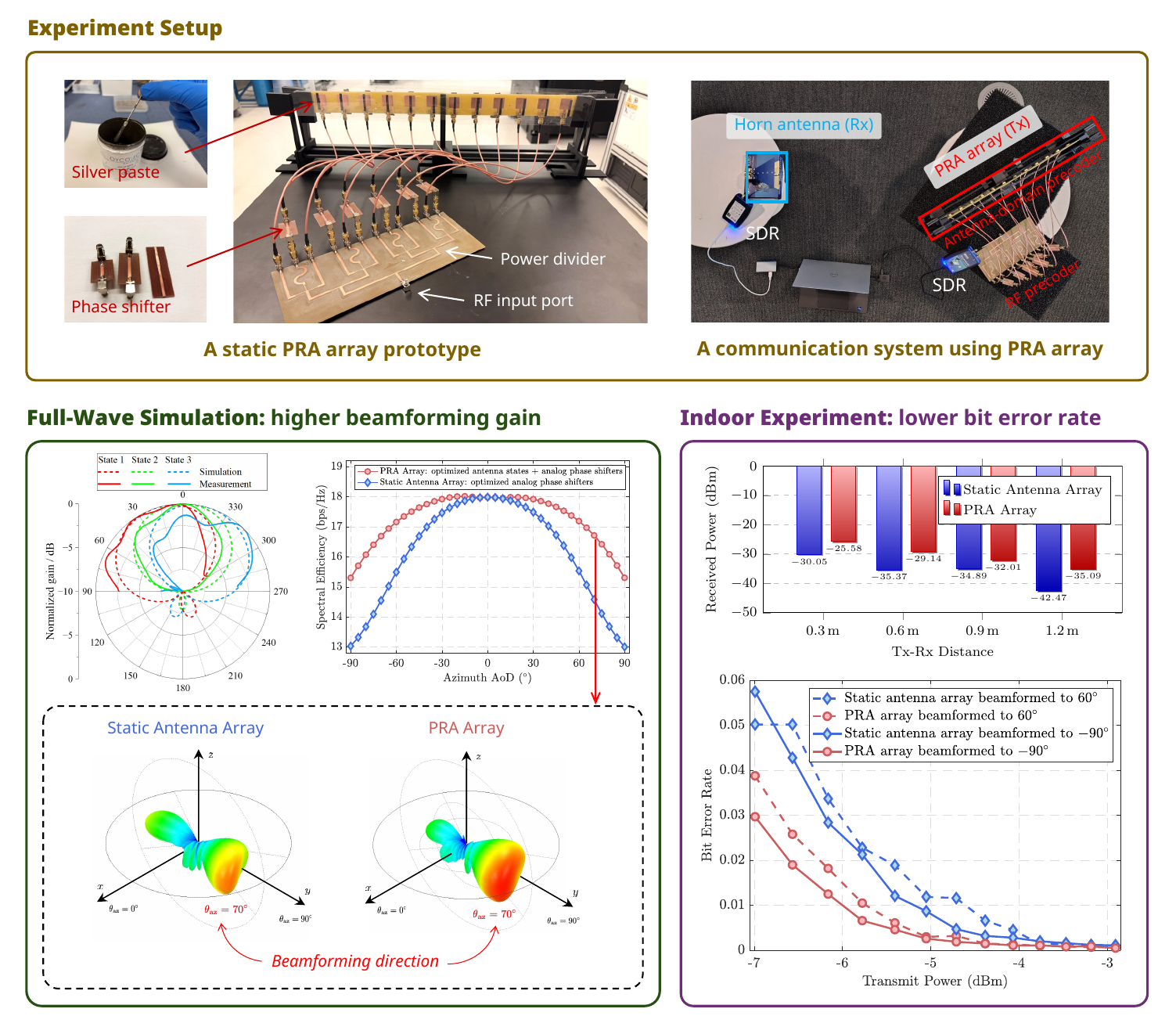}
\caption{
  {Experimental setup and results with the fabricated ERA communication system in~\cite{Wang2025Electromagnetically}. We compare the performance of the prototype with a static antenna array benchmark. The comparison covers both radiation performance (through full-wave simulations) and communication performance (through indoor experimental measurements).}
}
\label{fig_PZ2}
\end{figure*}

The experimental setup is illustrated at the top of Fig.~\ref{fig_PZ2}. The entire antenna array comprises 12 \ac{PRA} elements arranged in a uniform linear array. Notably, each antenna element has a physical size of $\unit[0.5]{\lambda} \times \unit[0.5]{\lambda}$, allowing for a densely packed half-wavelength-spaced array. The \ac{PRA} array is integrated with an analog feeding network and an ADALM-Pluto \ac{SDR} to form a complete communication transmitter. The feeding network consists of 12 \ac{RF} phase shifters, enabling analog beamforming, while the \ac{SDR} performs baseband signal processing. A standard SH2000 horn antenna paired with another ADALM-Pluto \ac{SDR} serves as the receiver, forming a point-to-point communication link. As shown in Fig.~\ref{fig_PZ2}, we test communication performance in an indoor environment.

We conduct performance evaluations through both full-wave \ac{EM} simulations and indoor experimental measurements. The results are summarized in Fig.~\ref{fig_PZ2}, where we compare the fabricated \ac{PRA} array against a benchmark static antenna array with fixed individual radiation patterns. {For a fair comparison, the \ac{PRA} array and the static benchmark employ the same 12-element aperture, half-wavelength element spacing, \ac{RF} input power, and analog feeding network, and thus incur the same feeding-network loss.} {The only difference is that each element of the \ac{PRA} array can dynamically select its radiation pattern from the three representative states shown in Fig.~\ref{fig_PZ2}, whereas the static benchmark fixes all elements to State~2.} {The full antenna prototype reported in~\cite{Wang2025Electromagnetically} supports 64 total radiation states, but only these three representative states are used here to illustrate the prototype comparison.} The communication experiments are conducted using 4-QAM modulation.

\begin{itemize}
  \item \textbf{Radiation Performance Evaluation:} As shown in the left column of Fig.~\ref{fig_PZ2}, the \ac{PRA} array exhibits a significant improvement in array gain compared to the benchmark static antenna array. The overall beampattern of the \ac{PRA} array is effectively concentrated in the desired direction without notable sidelobes, which demonstrates its superior beamforming capability.
  \item \textbf{Indoor Experimental Measurement:} As shown in the right column of Fig.~\ref{fig_PZ2}, the enhanced beamforming capability of the \ac{PRA} array results in a considerable improvement in received signal power during experiments. This subsequently leads to a significant reduction in communication \ac{BER}, particularly at low transmit power levels. {These results validate the capability of \acp{RA} to improve received signal power and link reliability in practical systems.}
\end{itemize}

\subsection{Limitations}

{While the experimental results demonstrate the strong potential of \acp{RA} to enhance communication performance, the current prototype should be viewed as exploratory rather than deployment-ready or optimal. It serves as a proof-of-concept case study illustrating how element-level reconfigurability, when jointly considered with array scalability, can be effectively leveraged to improve communication performance in practical transceiver implementations.}
Meanwhile, the fluid-metal-based actuation employed in the presented \ac{PRA} array suffers from slow reconfiguration speed and nontrivial fabrication and maintenance complexity. 
These limitations motivate the exploration of alternative actuation mechanisms, such as  microelectromechanical systems or high-speed electronic switches, as well as more robust and cost-efficient materials. Importantly, the observed communication gains are not tied to the specific fluid-metal implementation, but instead originate from the additional antenna-domain degrees of freedom enabled by element-level pattern reconfigurability. From a broader perspective, this prototype suggests that the practical value of RAs for communication systems lies in array-native designs with independently reconfigurable elements that can be jointly optimized.

Beyond these hardware considerations, fully leveraging the capabilities of \acp{RA} requires advanced system-level design and signal processing. The following sections explore promising architectural strategies and signal processing approaches to help bridge the gap between prototype demonstrations and practical deployment.

\section{Tri-Hybrid Architectures}

Effectively integrating the additional degrees of control offered by \acp{RA} into practical transceivers remains a critical challenge. The core consideration is to balance hardware cost, energy consumption, and communication performance. In this context, the tri-hybrid digital-analog-antenna architecture has recently emerged as a promising paradigm~\cite{Heath2025Tri,Castellanos2025Embracing,Deshpande2026Signal}.

\begin{figure*}[t]
  \centering
  \includegraphics[width=0.88\linewidth]{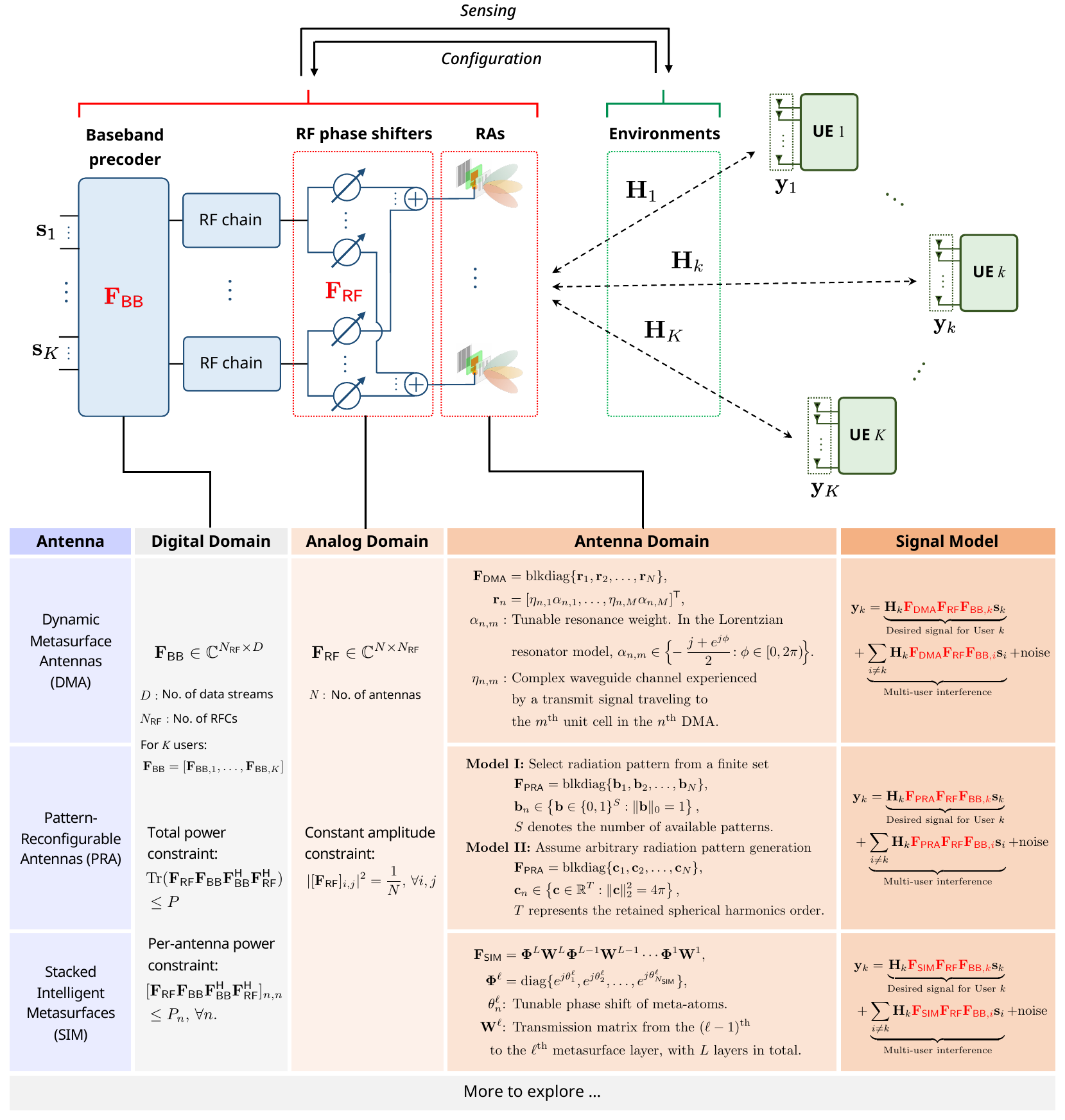}
  \caption{
  Illustration of a multiuser \ac{MIMO} communication scenario employing a tri-hybrid precoding architecture. The antenna domain precoding can be enabled by various types of \ac{RA} technologies. As representative examples, several antenna-domain models based on DMA~\cite{Castellanos2023Energy}, PRA~\cite{Zheng2025Tri}, and SIM~\cite{An2024Two} are presented.
}
  \label{fig_PZ3}
\end{figure*}

The tri-hybrid architecture is a wireless transceiver design that integrates processing across the digital (baseband), analog (\ac{RF}), and antenna (\ac{EM}) domains~\cite{Heath2025Tri}, as illustrated in Fig.~\ref{fig_PZ3}. It extends the conventional hybrid digital-analog architecture~\cite{Ayach2014Spatially, Sohrabi2017Hybrid}, incorporating the antenna domain as an additional processing space, rather than treating the antenna array as a static front-end. The tri-hybrid architecture holds practical significance for next-generation communication systems, wherein the digital domain enables multi-stream data processing, the analog domain mitigates hardware cost and power consumption, and the antenna domain introduces additional degrees of reconfigurability to effectively realize larger \ac{EM} apertures without proportionally increasing hardware and energy costs. {This system-level shift offers new performance-energy-cost tradeoffs that are unattainable with conventional hybrid designs, particularly in large-scale yet energy-constrained MIMO deployments~\cite{Heath2025Tri}.}

An interesting characteristic of the tri-hybrid architecture, as illustrated in Fig.~\ref{fig_PZ3}, is that the antenna-domain processing can be realized through various types of \ac{RA} technologies. Each antenna technology employs distinct models and provides unique performance benefits. For instance, tri-hybrid architectures incorporating \ac{DMA} exhibit superior energy efficiency~\cite{Heath2025Tri}, whereas implementations utilizing \ac{SIM} offer enhanced computational efficiency and improved signal direction detection capabilities~\cite{An2024Two}.

\section{Signal Processing Challenges}

{When \acp{RA} are integrated into communication systems, the resulting architectures introduce unique signal processing challenges.} In general, the associated challenges primarily lie in two aspects: (i) accurately sensing the wireless environment, and (ii) optimally configuring the antenna, analog, and digital domains in a tightly coupled manner.

\subsection{Wireless Environment Sensing}

{The sensing advantage of \acp{RA} lies not only in beamforming enhancement, but also in enabling additional \ac{EM} degrees of freedom for environment probing and observation. In conventional phased arrays, sensing is mainly performed through digital or analog beamforming over a fixed aperture structure. By contrast, \ac{RA}-enabled architectures allow the \ac{EM} response of the aperture itself to be reconfigured~\cite{Fang2026Tri, Fadakar2026Hybrid, Zhang2026Tri}. Different antenna states can therefore produce distinct effective array responses for the same physical aperture, creating additional measurement diversity that improves angular discrimination and scene observability without proportionally increasing the number of antenna elements or \acp{RF} chains. This capability is particularly relevant to emerging 6G sensing and positioning applications, where representative use cases already target meter-level and sub-meter-level accuracy~\cite{Wymeersch2025Cross}.

\subsection{System Configuration Optimization}

When the wireless channel environment is well sensed and characterized, the next critical step is to properly configure the system to maximize desired performance based on the acquired environmental information. In RA-based systems, this problem departs from conventional beamforming optimization by introducing antenna configuration as an additional and tightly coupled design dimension. To achieve optimal performance, the configuration of \acp{RA} must therefore be jointly optimized with other system components (e.g., digital and analog precoders/combiners).

\begin{figure*}[t]
  \centering
  \includegraphics[width=0.95\linewidth]{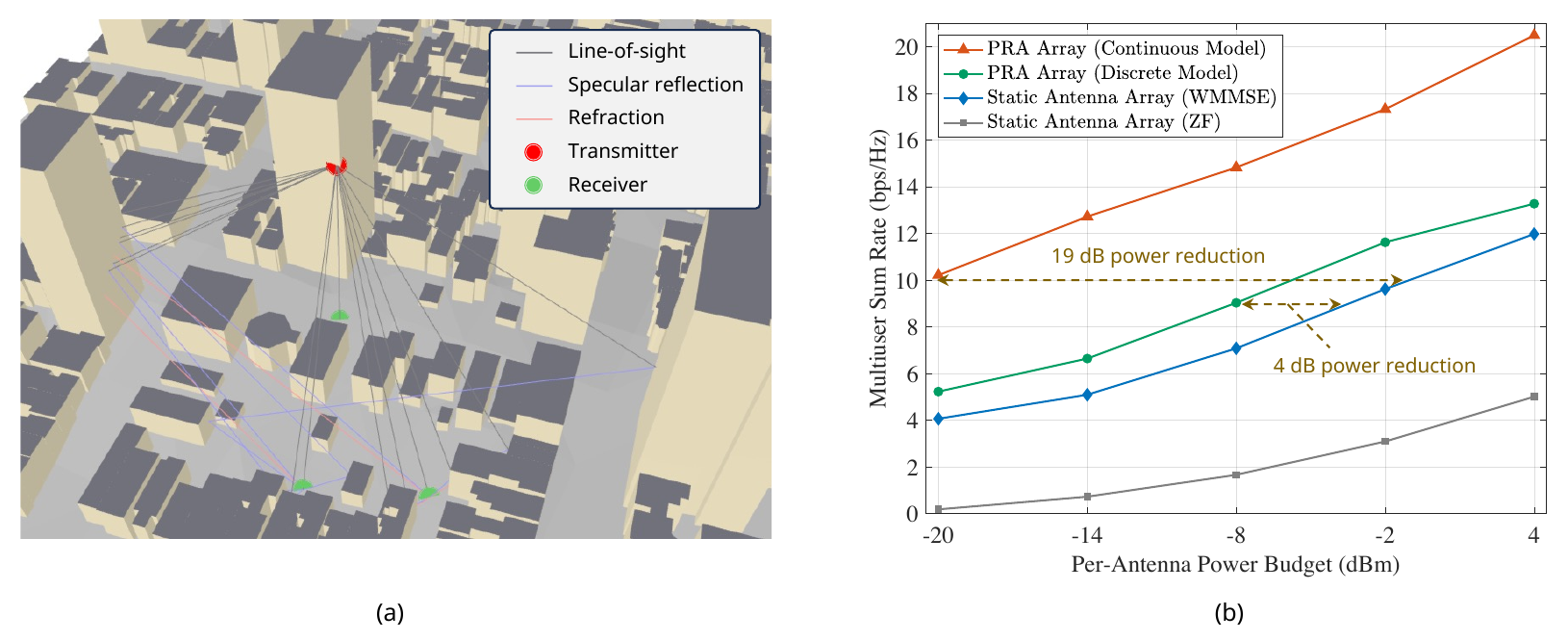}
    \caption{Evaluation of the multiuser system sum rate. (a)~Simulation scenario set in the San Francisco urban environment, with one transmitter equipped with a $10\times10$ antenna array and three users, each equipped with a $2\times2$ antenna array. The ray-tracing simulation is conducted using Sionna~RT~\cite{sionna}. (b)~System sum rate versus the per-antenna power budget at the transmitter under different antenna models and optimization methods.}
  \label{fig_PZ5}
\end{figure*} 

As an illustrative example, Fig.~\ref{fig_PZ5} presents an evaluation of the system sum rate versus the per-antenna power budget at the transmitter for a multiuser \ac{MIMO} system. 
In this study, two models for \ac{PRA} operation are considered: 
\begin{itemize}
    \item \textbf{Discrete Model:} each antenna element selects its radiation pattern from a set of 64 predefined candidates, which are measured from the \ac{PRA} prototype reported in~\cite{Wang2025Electromagnetically}.
    \item \textbf{Continuous Model:} each antenna element can generate arbitrary radiation patterns. {This is an idealized upper-bound model used to assess the performance potential of highly flexible antenna-domain control.}
\end{itemize} 
We compare the \ac{PRA} array against the static antenna array. 
For the static array, the beamforming configuration is optimized using (i) the \ac{WMMSE} algorithm~\cite{Shi2011An} and (ii) the \ac{ZF} algorithm~\cite{Spencer2004Zero}, whereas for the \ac{PRA} array, both beamforming and antenna configuration are jointly optimized using the algorithm proposed in~\cite{Zheng2025Tri}.

The results show that the employed algorithm effectively configures the \ac{PRA}-equipped system to achieve high sum rates, consistently outperforming conventional static arrays. 
{Essentially, these gains stem from the ability of \acp{PRA} to actively reconfigure the wireless channel itself, rather than solely from digital-analog beamforming, which is the only optimization lever available in static arrays.}
{Moreover, the \ac{PRA} array can achieve comparable system sum rates with lower transmit power in this simulated setup, demonstrating its potential for system energy savings.}
{Finally, \acp{PRA} with arbitrary radiation patterns (continuous model) substantially outperform their discrete-model counterparts based on current hardware prototypes. The continuous model is interpreted only as an idealized upper bound, since constraints on radiation efficiency, bandwidth, passivity, and realizable current distributions are not imposed. Accordingly, the gap between the continuous and discrete models should be viewed as the performance headroom between current hardware prototypes and an idealized limit, rather than as a directly achievable practical gain.}

Although several static optimization methods for reconfigurable antenna configurations have been proposed, a more critical and RA-specific challenge lies in the development of real-time and adaptive optimization techniques that can track both channel dynamics and antenna reconfiguration states. To date, this problem has not been thoroughly studied.

\subsection{AI-Based Solutions}

{AI has emerged as a promising tool for addressing the above challenges in \ac{RA} systems. Existing studies along this line mainly follow two distinct research directions:}
\begin{itemize}
\item {\textbf{AI for RA Configuration:} Since the use of \acp{RA} leads to configuration-dependent channel responses and tightly coupled multi-domain optimization, which challenge conventional model-based signal processing, learning-based approaches have emerged as an effective solution~\cite{Zheng2026Fast}. In particular, well-trained deep neural networks can directly infer proper antenna, RF, and baseband configurations from observations without explicit channel estimation~\cite{Sohrabi2022Active}.}
\item {\textbf{\ac{EM} Hardware for Neural-Network-Like Computation:} Beyond using AI to configure \acp{RA}, recent studies further show that \acp{RA} and related programmable \ac{EM} structures can physically implement neural-network-like processing in hardware-in-the-loop (HIL) settings~\cite{liu2022programmable,Gao2023Programmable,Govind2025Integrated}. In this case, the \ac{EM} hardware itself participates in inference or computation, potentially enabling improved hardware efficiency.}
\end{itemize}

\section{Conclusion}
RAs are becoming a key enabler for next-generation wireless communications, as their adaptability can significantly improve energy and hardware efficiency, while compact and scalable antenna designs support large-scale deployments. However, realizing their full potential requires tighter integration of practical RA hardware with new transceiver architectures and advanced signal processing. Future research directions include novel RA designs for 6G, the development of efficient and adaptive control algorithms, and system-level prototyping and experimental validation.

\bibliography{references}
\bibliographystyle{IEEEtran}

\end{document}